\documentclass[11pt,a4paper]{article}%
\usepackage{amsfonts}
\usepackage{graphicx}
\usepackage{amsmath}
\usepackage{hyperref}
\usepackage{enumerate}%
\usepackage{amssymb}
\providecommand{\U}[1]{\protect\rule{.1in}{.1in}}
\makeatletter \@addtoreset{equation}{section}

\newcommand{\be}{\begin{equation}}
\newcommand{\ee}{\end{equation}}
\newcommand{\bea}{\begin{eqnarray}}
\newcommand{\eea}{\end{eqnarray}}
\begin{document}

\title{{ \begin{flushright}
{\normalsize\small CPM-11-08/ LPHE-MS-11-08 }
\end{flushright} }\textbf{\ On    Fractional Quantum Hall Solitons}\\{ \bf   in   ABJM-like
Theory} }
\author{Adil  Belhaj\thanks{belhaj@unizar.es}, \hspace*{-15pt}\\
\\{\small Centre of Physics and Mathematics, CPM-CNESTEN, Rabat, Morocco}
\\{\small Lab Phys Hautes Energies, Modelisation et Simulation, Facult\'{e}
des Sciences, Rabat, Morocco}
\\ {\small
Groupement National de Physique des Hautes Energies, Si\`{e}ge
focal: FSR, Rabat, Morocco }} \maketitle

\begin{abstract}
Using D-brane physics, we  study  fractional quantum Hall solitons
(FQHS) in ABJM-like theory  in terms of    type IIA dual geometries.
In particular,   we discuss a class of Chern-Simons (CS) quivers
describing  FQHS sytems  at low energy. These CS quivers come from
R-R gauge  fields  interacting with D6-branes wrapped on 4-cycles,
which reside within a blown up $\bf {CP}^3$ projective space. Based
on  the  CS quiver method  and mimicking the construction of del
Pezzo surfaces in terms  of $\bf CP^2$, we first give a model which
corresponds  to a   single layer model of FQHS system, then we
propose a multi-layer system generalizing the doubled CS field
theory,  which is  used in the study of topological defect in
graphene.
\end{abstract}

\newpage

\section{Introduction}
There have  been considerable works dealing with   connections
between string theory and  quantum Hall solitons (QHS) in
(1+2)-dimensions. The first of them was a construction of Bernevig,
Brodie, Susskind and Toumbas describing such models   on a 2-sphere
using a spherical D2-brane and dissolved $N$ D0-branes moving on
it\cite{BBST}. The system has been placed in a background of
coincident D6-branes extended in the directions perpendicular to the
world-volume of the D2-brane on which the quantum  Hall  effect  (QHE)
resides.  The    ten dimensional string picture of   the  QHE has
been extended to the compactification of type IIA superstring theory
on the  K3 surface with  singularities classified by Lie algebras.
This extension  has been based on the study of quiver gauge theories
living on the world-volume of the wrapped D4-branes on intersecting
2-spheres arranged as  Dynkin Diagrams\cite{BS,BEFKSS}.
\par Recently, some efforts have been devoted to study fractional
quantum Hall solitons(FQHS) in relations with in Anti de
Sitter/conformal field theory (AdS/CFT)
correspondence\cite{FLRT,BJLL}. In particular,
 connections  with (2+1)-dimensional Chern-Simons (CS) theory
constructed by Aharony-Bergman-Jafferis-Maldacena (ABJM) theory have
been discussed in \cite{Fujita,QHEinABJM}.   The latter is a
3-dimensional
  $N=6$ CS  quiver  with  $\mbox{U(N)}_k \times  \mbox{U(N)}_{-k}$  gauge symmetry  proposed to be dual to M-theory
  propagating on $AdS_4\times S^7/Z_k$, with an appropriate amount of fluxes, or type IIA superstring on
    $AdS_4\times \bf CP^3$ for large number of $k,N$ with $k\ge N$ in the weakly interacting regime  \cite{ABJM}.
    In the decoupling limit, the corresponding  CFT$_3$  is generated by
the action of multiple M2-branes placed at the orbifold $
C^4/Z_k$\footnote{ see also \cite{rey}.}. In this regard,  it has
been shown that QHE can be obtained from the world-volume action of
the M5-brane filling AdS$_3$ inside AdS$_4$\cite{Fujita}. The model
has been derived from  $d = 3$ flavored ABJM theory with the CS
levels $(1,-1)$. Alternatively,   a FQHE    system   in
AdS$_4$/CFT$_3$ has been realized by adding fractional D-branes
(D4-branes wrapping $\bf {CP}^1$) to the ABJM
theory\cite{QHEinABJM}.

Motivated by these investigations, we contribute to these activities
by discussing  FQHS  in AdS$_4$/CFT$_3$
 using  type  IIA dual geometry realized as blown up of $\bf
 {CP}^3$ by four-cycles which are isomorphic  to  $\bf  {CP}^2$. In toric
geometry,  this procedure  extends  the  building  of  del Pezzo
surafces from $\bf  {CP}^2$ to  $\bf  {CP}^3$.   We  will
     refer   to the corresponding CS quiver
     theory  as  ABJM-like theory. In  this work,   we
give  a class of  such  quivers describing FQHS from D6-branes
interacting with  R-R gauge fields in    ABJM-like geometry. We
propose  a stringy hierarchical description in terms of wrapped
D6-branes on  the blown up  four-cycles. We first consider   a model
corresponding to a   single layer of  FQHS, then  we discuss a
multi-layer system  which generalizes  the doubled CS  field theory
which is  used recently in the  topological defect  in
graphene\cite{graphene}.\par
\section{U(1)$_k$ Chern-Simons  theory and ABJM  theory}
 To start, recall that   fractional quantum Hall  states  were proposed first
by Laughlin and they are characterized by the filling factor
$\nu_{L}=\frac{1}{k}$ where $k$ is an even integer for a boson and
an odd integer for a fermion\cite{L}. At low energy, this model can
be described by a 3-dimensional U(1)$_k$ Chern-Simons  theory. The
corresponding effective action reads  as
\begin{equation}
S_{CS}=-\frac{k}{4\pi}\int_{\mathbb{R}^{1,2}}A\wedge
dA+\frac{q}{2\pi}{\tilde{A}}\wedge dA\label{sc},
\end{equation}
where $A$ is the dynamical gauge field, $\tilde{A}$  is  an external
electromagnetic field, and $q$ is the charge of the electron
\cite{wen,WZ}. In the present work, we will see that   U(1)$_k$  CS
theory, describing FQH, can be derived  from ABJM theory  with
$\mbox{U}(N)_k\times \mbox{U}(N)_{-k}$ gauge symmetry.
     Roughly speaking,  U(1)$_k$ CS gauge theory  can be obtained from
D6-branes   wrapped   on four-cycles   which are  embedded in type
IIA geometry.   Indeed, let us take a stack of $M$ D6-branes in
AdS$_4\times  \bf CP^3$ type IIA geometry.  A priori, D6-branes can
wrap various cycles. However, here, we consider a stack of $M$
D6-branes  wrapping a particular  complex hypersurface class $[C]$
in $H^4( {\bf CP^3},Z)$ which is one dimension.  On the gauge
theory side, the gauge symmetry $\mbox{U}(N)_k\times
\mbox{U}(N)_{-k}$ becomes $\mbox{U}(N+M)_k\times \mbox{U}(N)_{-k}$.
 To get the
action (\ref{sc}),  we take just the U(1)  part of $\mbox{U}(M)_{k}$
corresponding to a single   D6-brane wrapping  the four-cycle $C$
and consider other gauge factors   as spectators as  made in
\cite{Fujita,QHEinABJM}. Indeed, on a seven dimensional world-volume
of the D6-brane lives an U(1) gauge symmetry with the following
action
\begin{equation}
S_{D6}=S_{DBI}+S_{WZ}
\end{equation}
where
\begin{equation}
S_{DBI}\sim T_{6} \int_{\mathbb{R}^{1,6}}  e^{-\phi}\sqrt{
-\det(G+2\pi F)}
\end{equation}
and where the action  $S_{WZ}$ depends on the R-R gauge fields of
type IIA superstring. To obtain  the first
 term of Chern-Simons action, we  ignore  the $S_{DBI}$  action and take
 $S_{WZ}$  as
\begin{equation}
S_{WZ}\sim T_{6}\int_{\mathbb{R}^{1,6}} F\wedge F\wedge A_{3},
\end{equation}
 where $T_{6}$ is the D6-brane tension and where $A_{3}$ is the R-R
3-form coupled to the D2-brane of type IIA superstring.   Performing
a simple integration by parts and  integrating  the result  over
$C_4$, we get the first term of the action  (\ref{sc}), namely
\begin{equation}
-\frac{k}{4\pi} \int_{\mathbb{R}^{1,2}} A\wedge F
\end{equation}
where $k=\frac{1}{2\pi}\int_{C_4}(dA_3)$  is  produced  now  by   $k
$ D4-flux. To couple the system to an external gauge field, we need
to turn on the RR 5-form $A_{5}$ which is coupled to the D4-brane.
This gauge field  should be decomposed as follows
\begin{equation}
A_{5}\to\tilde A \wedge\omega
\end{equation}
where $\omega$ is a harmonic 4-form  dual to  the four-cycle $C_4$.
In this way, the WZ term $\int_{\mathbb{R}^{1,6}} A_{5} \wedge F$ on
a D6-brane gives the second  term of the action (\ref{sc}), namely
\begin{equation}
q \int_{\mathbb{R}^{1,2}} \tilde A \wedge F
\end{equation}  where    $q=\int_{C}w$ and  $\tilde A$  is an  U(1)
gauge field  serves as the external gauge field that couples to
gauge fields living on  the D6-brane world-volume. Integrating out
the gauge field $A$ and using the electric current expression, the  above
effective action reproduces now  the following filling factor
\begin{equation}
\label{nu}
\nu=\frac{q^2}{k}.
\end{equation}
From this equation, it follows  that the filling factor depends on
the  D4-branes and  the  harmonic 4-forms  defined on $\bf CP^3$. It
turns out that  known values could  be reached by taking particular
choices of such parameters.
\section{FQH systems  with $\mbox{U}(1)\times
  \mbox{U}(1)$ gauge symmetry}
Once given a  system  with a  single U(1) gauge symmetry,  we will
discuss  a possible generalization  described by an  effective
Chern-Simons gauge theory with a series of U(1) gauge fields. The
general form of the   abelian part  of these effective theories is
given  by
\begin{equation}%
\begin{tabular}
[c]{ll}%
$S\sim\frac{1}{4\pi}\int_{\mathbb{R}^{1,2}} K_{ij}A^{i}\wedge
dA^{j}+2 q_{i}\tilde{A}\wedge dA^{i},$ &
\end{tabular}
\label{hd}%
\end{equation}
where  $K_{ij}$ is a real, symmetric and invertible matrix ($\det
K\neq 0 $)  with $q_{i}$  is a vector of charges.   The external
gauge field  $\tilde{A}$
couples  now to each   current $\star dA^i$ with charge strengths  $eq_i$. The $K_{ij%
}$\ matrix and the $q_{i}$  charge vector in this effective field
action are  suggestive of some physical concepts. Following the
Wen-Zee model \cite{WZ}, $K_{ij}$ and $q_{i}$ are interpreted as
order parameters and  classify the various QHS states. Integrating
over the all  gauge fields $A^{i}$, one gets  the formulae for the
filling factor
\begin{equation}%
\begin{tabular}
[c]{ll}%
$\nu=q_{i}K_{ij}^{-1}q_{j}.$ &
\end{tabular}
\label{factor}%
\end{equation}
In the following,    we will see that  the corresponding    models
can be also obtained from D6-branes wrapping individually more than
one four-cycles in type IIA geometry. In ABJM backgrounds,  one way
to achieve this  extended geometry is  to use  the  blown up  $\bf CP^3$ at
singular toric vertices. In particular, we will  explore  the same
analysis  for constructing del Pezzo surfaces from the projective
space $\bf CP^2$ using toric geometry language\cite{F,LV}. Indeed,
the meaning of  a blown up $\bf CP^2$
 is to replace a point on it  by a $\bf CP^1$. In  toric geometry framework,
  the  realization of $\bf CP^2$ involves a
triangle  over each point of which there is a 2-torus $T^2$. Blowing
up $\bf CP^2$ consists of replacing a  vertex of  the  triangle
by a $\bf CP^1$ and that is realized in toric geometry by an
interval. This geometric operation  can be extended to   $\bf CP^n$
which is  identified with $T^n$  fibered  over a  $n$-dimensional
 simplex.   For $n=3$, the  toric realization is given   by  a
 tetrahedron.
Applying the blowing up procedure to $\bf CP^3$, we can replace one
vertex of the tetrahedron by  an exceptional four-cycle $E$ which is
isomorphic to $\bf CP^2$. In toric geometry, this  is realized by a
triangle. The resulting blown up space $\tilde{\bf CP}^3$ has now
two Kahler parameters and the 4-cycle homology group $H^4( {\tilde
{\bf CP^3},Z)}$ is two dimension and is generated by $\{C_4,E\}$.
Roughly speaking,  consider  now  type IIA superstring on
AdS$_4\times \tilde {\bf CP}^3$ in the presence of a new stack of
$M'$ D6-branes wrapping  on the  blown up $E$ space. Inspired by the
study of quiver gauge theories in connection with AdS/CFT
conjecture, this blowing  up can be understood by  the introduction
of a  new gauge factor. In particular, the appearance of the new
factor is partly motivated by the result of the geometry of
$AdS_5\times M_5$, where $M_5=S^5/\Gamma$ with $\Gamma$ being a
discrete subgroup of SO(6),   dual  to D3-branes at ADE
singularities of Calabi-Yau threefolds generalizing the conifold
singularity \cite{KS,GNS}. It can be also   supported  also by
results on $N = 1$ four-dimensional quiver theories arising on the
world-volume of D3-branes transverse to cones over complex
two-dimensional toric varieties, e.g. del Pezzo surfaces \cite{HI}.
Based on  such quivers built  on $\bf CP^2$ and  del Pezzo surfaces,
we propose that  new gauge factors  can be enhanced due to the
blowing up procedure. In this way of  thinking,  the gauge symmetry
of ABJM $\mbox{U}(N)_k\times \mbox{U}(N)_{-k}$ changes into
$\mbox{U}(N+M)_k\times \mbox{U}(N)_{-k} \times \mbox{U}(M')_{k'}$.
The corresponding CS  theory is represented by a quiver
with  three vertices which  is represented by the following triangle
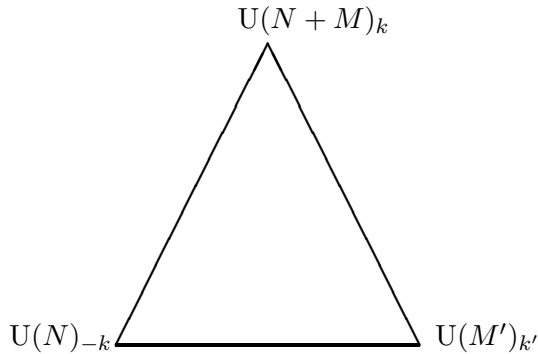
\begin{equation*}
\mbox{
  \begin{picture}(100,132)(0,0)
  \unitlength=2cm
  \thicklines
   \put(0,0){\line(1,2){1}}
   \put(0,0){\line(1,0){2}}
   \put(2,0){\line(-1,2){1}}
   \put(0.8,2.1){$ \mbox{U}(N+M)_k$}
   \put(-0.7,0){$\mbox{U}(N)_{-k}$}
   \put(2.1,0){$\mbox{U}(M')_{k'}$}
 \end{picture}
  }  \label{four}
\end{equation*}
 We will not study the general case. Instead, we will discuss a particular model
 motivated by the study of topological defect  in  graphene. This  model  which is   called  doubled level-$k$
  CS field theory  $\mbox{U}(1)\times
  \mbox{U}(1)$  is
dealt with  in \cite{graphene}. The  action of this model takes the
form  \begin{equation}
S_{CS}=-\frac{1}{4\pi}\int_{\mathbb{R}^{1,2}}kA^1\wedge
dA^1-kA^2\wedge dA^2. \label{doubled}
\end{equation}
To make   contact  with IIA superstring on AdS$_4\times \tilde
{CP}^3$, one
 requires  the following  condition on CS
levels and gauge group ranks
\begin{equation}
k(M+N)-kN+k'M'=0.
\label{constraint}
\end{equation}
This is  inspired  from the study of  CS  quivers  with $
\prod_i\mbox{U}(N_i)_{k_i}$ gauge symmetry\cite{MS},  where the
constraint on CS  levels  $k_i$ is given by
\begin{equation}
\label{cong} \sum_ik_iN_i=0.
\end{equation}
Taking into account this observation, the link  we are after force
us to consider  a particular   solution given by
\begin{equation}
M=M'\qquad k'=-k.
\end{equation}
As before, thinking that  ABJM symmetry as spectators, the CS quiver
theory
 describing  FQH system   will be in the   $\mbox{U}(M)_k\times \mbox{U}(M)_{-k}$ part
of $\mbox{U}(N+M)_k\times \mbox{U}(N)_{-k} \times \mbox{U}(M)_{-k}$.
Extracting an  $\mbox{U}(1)\times \mbox{U}(1)$ abelian part of
$\mbox{U}(M)_k\times \mbox{U}(M)_{-k}$,  we  can obtain  the
doubled  CS field theory model  from  two D6-branes wrapping
individually  the two four-cycles $C_4$ and $E$.  Using  quiver  gauge
theory living on their world-volumes and proceeding  now similarly
as in the preceding model,   we can  get a model  with the following
 $K_{ij}$ matrix
\begin{equation}
K_{ij}=\left(
\begin{array}{cc}
 k & 0\cr
0&-k
\end{array}%
\right). \label{hd1}
\end{equation}
 This matrix, which  describes  a system without interactions between gauge
fields,  produces  the effective action given in (\ref{doubled}).
Indeed, the first term of this action can be obtained from a
D6-brane wrapping $C_4$ as we did before. Similarly, the  second
term can be derived from D6-branes wrapping $E$ in the presence of
the R-R gauge field sourced by the anti-D4-flux,   which leads to
$-k=\frac{1}{2\pi}\int_{E}(dA_3)$. To couple the system to an
external gauge field, one needs an extra D4-brane coupled to a RR
5-form $A_{5}$. This gauge field  can be decomposed in terms of a
harmonic 4-form which is   dual to a   generic 4-cycle described by
a linear combination of $\{C_4,E\}$ as follows
\begin{equation}
C=q_1C_4+q_2E.
\end{equation}
Evaluating (\ref{factor}) for the charges $q_i = (q_1, q_2)$ yields
\begin{equation}
\nu=\frac{q_1^2-q_2^2}{k}. \label{factorb}
\end{equation}
Let us  give a concrete  example to illustrate  how the formulae
works. Taking  $q_1=2$, $q_2=1$  and  specializing the general
expression  to  the value $k=2$, we obtain $\nu=\frac{3}{2}$. The
corresponding   model   has been studied  extensively theoretically
and experimentally. The computation for $q_1=n+1$, $q_2=n$ and
$k=n+1$, gives a remarkable sequence $\nu=\frac{2n+1}{n+1}$. In this
way, we can generate  the following values
\begin{equation}
\frac{3}{2}\to \frac{5}{3}\to \frac{7}{4} \to \frac{9}{5}\ldots
\label{sequence}
\end{equation}
It is worth noting at this point  that  this  sequence   can be
viewed as the inverse of  Fary series  used in the  hierarchy
structure and scaling theory of FQHE \cite{sebbar}. It follows that,
the vanishing filling factor behavior has to do with the line
defined by $q^2=0$ in the $(k, q^2)$ parameter space, where
$q^2=q_1^2-q_2^2$. This can  happened  if the condition $q_1=q_2$ is
satisfied, which
 means that we can fix the world-volume flux by the number
of D4-branes.  The Hall conductivity is quantized in terms of the CS
level being identified with the D4-flux  and the quadratic charge
$q^2$ given in terms of the integral of  harmonic 4-forms over   dual four-cycles used
in the the blowing up of  $\bf CP^3$.
\section{On  extended models}
 In this section, we would like to note that the above
analysis can be extended in many ways. On of them is to use  the
blown up $\bf CP^3$ at $\ell$ points. In this way, the 4-cycle
homology group $H^4(\tilde CP^3,Z)$ is now  $\ell +1$ dimension
 and is generated by $\{C_4,E_\ell\}$.  In type IIA superstring theory, the
corresponding   model appears as  a CS quiver model on  the
world-volume of $\ell +1$ D6-branes wrapped separately on  4-cycles
generating  $H^4(\tilde CP^3,Z)$. On the world-volume of theses
wrapped D6-branes lives an U(1)$^{\ell+1}$ gauge  theory in
3-dimensional space-time on which FQHS system will reside. The
general study is beyond the scope of the present work, though we
will consider a simple brane system extending the  above doubled CS
field theory model. This can be obtained by requiring that the
matrix CS coupling to be
\begin{equation}
K_{ij}=k\eta_{ij}\label{kij}
\end{equation}
where  $\eta_{ij}=diag(\ell,-1,\ldots,-1)$. This matrix  is obtained
by solving a similar
 equation  as  the one appearing in  (\ref{constraint}).  As in the previous
 discussed model, the   external gauge field arises from the decomposition of  the  RR 5-form $A_{5}$  in
 terms of a comabination of harmonic four-forms where the dual four-cycle takes the following form
\begin{equation}
C=q_1C_4+q_2E_1+\ldots+ q_{\ell+1}E_{\ell}.
\end{equation}
  For the vector charge
$q_i = (q_1, \ldots, q_{\ell+1})$, the filling factor  can be
simplified  as
\begin{equation}
\nu=\frac{q^2}{\ell k} \label{factorb}
\end{equation}
where  $q^2=q_1^2-\ell\sum\limits_{i=1}^{\ell+1} q_i^2$.  For
$\ell=1$,   we recover the model associated with  doubled CS  field
theory.  As before, the vanishing filling factor behavior has to do
with the line defined by $q^2=0$ in the $(k, q^2,\ell)$ parameter
space. This can happened if the condition
$q^2_1=\ell\sum\limits_{i=1}^{\ell+1} q_i^2$ is satisfied  which
means that we can fix the world-volume flux by the number of
D4-branes  and the number of the  blown  up points.
\section{Discussion}
In this paper, we have  discussed   FQHS in ABJM-like theory.  The corresponding geometries are realized
 as  a  blown up $\bf CP^3$  projective space. To reach such systems, we have  considered
 abelian parts and  we have  assumed
the  remaining  part as spectators. Using toric geometry, we have
built CS quivers describing   FQHS  in terms of D6-branes and R-R
gauge fields. For the  blown  up  $\bf CP^3$  at $\ell$ points, we
have  given  a model  where the filling factor depends on the extra 
parameter $\ell$.  It is  worth  noting that  this  new parameter
$\ell$ gives more
  freedom to recover some known values of  the  filling factor.
 It follows also that  as we  blow up the type IIA geometry, the
FQHS system
 undergoes a scaling transformation   from  $\nu=\frac{q^2}{k}$ to $\nu=\frac{q^2}{\ell k}$. It should be
 interesting to understand, in more detail, this  scaling of the  CS couplings in terms of Khaler moduli
 space of type IIA geometry.  \par This work comes up with many open
 questions. First, we have analyzed a blown  up  $\bf CP^3$, it
 would  be important to consider  3-dimensional toric
 varieties extending   $\bf CP^3$  using technics of F-theory compactifications.    On the other
hand, it has been given a nice string
 realization of graphene \cite{rey1}  and its relation to CFT$_3$,  it  should be
 interesting  to make contact with  such a  3-dimensional   model. We
shall address these open questions in the future.

\hspace*{-15pt}

{\bf \emph{Acknowledgments}}:    The author thanks   M.  Asorey,
L.J. Boya, P. Diaz,  N. Fahssi, M.P. Garcia del Moral,  E.H. Saidi
and A. Segui for discussion,  collaborations on related topics and
scientific help. It is pleasure to thank  also Soo-Jong Rey   for
discussion and  scientific help.  He thanks also his
mother Haja Fatima    for  patience and support. This work is
partially supported by A/024147/09 and A9335/10.

\end{document}